# Leakage Localisation Using Cross Correlation for Wireless Three-Axis Vibration Sensors

**Airull Azizi Awang Lah [1], Rudzidatul Akmam Dzyauddin[2], Nelidya Md Yusoff [3], Mohd Ismifaizul Mohd Ismail[4], Noor Azurati Ahmad@Salleh[5], Robiah Ahmad[6]**

Razak Faculty of Technology and Informatics, Universiti Teknologi Malaysia Kuala Lumpur, Jalan Sultan Yahya Petra, 54100 Kuala Lumpur, Malaysia,
[1]aazizi9@graduate.utm.my,[2]rudzidatul.kl@utm.my, [3]nelidya.kl@utm.my, [4]ismifaizul@gmail.com, [5]azurati@utm.my, [6]robiahahmad@utm.my

**ABSTRACT**

The objective of this paper is to investigate the application of the cross-correlation technique in localizing the leakage in Acrylonitrile Butadiene Styrene (ABS) pipeline using wireless three-axis vibration sensors. Most of the existing leak localization techniques that utilized the cross-correlation method is using acoustic and pressure sensors. This study explored the impact of using three-axis vibration sensor ADXL335 to localize the leakages in ABS pipeline. The study focused on the small leak size of 1 mm and the effect of sensor distance and different water pressure is studied. The vibration data used in the cross-correlation data were filtered against the noise interference from the water pipeline. The source of noise interference from the water pipeline are motor, water pump, and valve. The computed leak localization accuracy against the actual leak distance from the three-axis was analyzed. The leakage localization formula is improvised by adding buffer time to have an accurate result. From the experiment, it was discovered that three axis vibration sensor can be used to localize the leakages in plastic water pipeline and it is suitable for deployment in small to medium water distribution network or within a building.

## 1. INTRODUCTION

Water is one of the basic and important matters in human life. Almost 65% of the human body contains water and it flows through blood, carrying oxygen and nutrients to cells and flushing out wastes from bodies. It also cushions the joints and soft tissues and without water as a routine part of our intake, we cannot digest or absorb foods [1]. On broader perspectives, water covers almost 75% on the earth surface. Though with abundance of water, 97.5% is salt water which cannot be consumed directly by a human. The remaining 2.5% is fresh water but apart from that only 0.007% is accessible to be consumed by 6.8 billion people all over the world [2]. To make matter worst, poor economics and bad infrastructure have caused millions of peoples especially children died almost every day from diseases affiliated by the scarcity of fresh water supplies, sanitation, and hygiene. More than 40% of the global population facing a risk of freshwater scarcity and this figure is forecasted to increase on yearly basis. According to United Nations, there are more than 783 millions of people lacking access to clean water and over 170 billion of people are living in an area where the water usage outstrips the supply [3]. Statistics from World Bank in 2016 summarizes that there 45 million cubic meters of water losses. This is equivalent to the loss of USD 3 billion and can supply water to 90 million people [4].

Water losses are one of the main challenges encountered by all water utility companies around the world. This losses related to the Non-Revenue Water (NRW), measured in percentage is an indicator of freshwater distribution efficiency in a geographical location. According to the International Water Association (IWA), NRW is the lowest hanging fruits to improve water utility efficiency [5]. The recommended NRW percentage outlined by the World Bank is 26% [6]. There are two main factors contributing to the NRW efficiency; physical losses and commercial losses. Physical losses caused by pipe burst, pipe leakage, and water overflow while the commercial losses were due to meter inaccuracies, illegal tapping, water maintenance and others.

The challenges in pipeline leakage detection and localization lies in the water distribution network itself, parameters estimation, non-linear nature of water network, users and the surrounding environments [7-8]. The current practice adopted by the water utility companies is to find the leak after received a report from the customer and the use of leak detection tools such as acoustic and pressure based sensors. The acoustic sensor was widely used in the industry to detect and locate the leak and one of it is Sahara system [9]. Although it is widely used in the industry, however according to [10] the acoustic method has a limitation in the non-metallic pipeline due to viscoelastic effect where the sound noise generated by the leak is absorbed by the material. This limitation has inspired other researchers to use vibration sensor to detect [11-13] and localize the leak [14-15]. In fact, the field of leak localization is relatively new compared to leak detection and most of the experiments and studies were conducted using single-axis accelerometer. [13] filled up the gap by exploring the use of three-axis accelerometer to analyze the leak detection and leak size.

## 2. EXISTING LEAK LOCALIZATION TECHNIQUES

The existing leak localization techniques can be divided into two broad categories; the external system and internal system. The external system can further be divided into two groups; the sensor based and non-sensor based. It is the use of external hardware either sensor based or non-sensor based mounted on or along the water pipeline to captured the required data for processing and generated an indicator. The sensors that used to localize the leaks are pressure sensor [15], wireless sensor networks (WSN) [16], mobile wireless sensor networks[17], acoustic sensor [9], Radio Frequency Identification (RFID) and WSN [18] and vibration sensor [13]. For non-sensor based, the hardware that has been used and explored are Optical Frequency Domain Reflectometry (ODFR) [19], Fiber Bragg Grating (FBG) [20], Ground Penetrating Radar (GPR) [21], electrokinetic [22] and Distributed Temperature Sensing (DTS) [23].

The internal system is the application of methods and techniques such as residuals, classifiers, algorithm, sensor placement method and others to process and analyze the required data from the pipeline. It can further be divided into two groups; sensor related and non-sensor related. The localization technique that based on sensor are digital recognition from dual sensors [24], Genetic Algorithm (GA) [25], multilabel classification [26], interval estimation [27], exhaustive search and GA [28] and sensor placement strategy [7, 29]. The localization technique that is non-sensor related are statistical classifier [8], Bayesian classifier[30], mixed-model and data driven approach [31], Kantorovich distance [32], Least Square Support Vector Machine (LS-SVM)[33], data driven and sensitivity analysis [34], differential algorithm[29], Signal Noise Ratio Empirical Mode Decomposition (SNR-EMD)[35], 2 stages approach [36] and Fisher Discriminant Analysis (FDA) [37].

This paper focused on using three-axis vibration sensor to localize the leak in ABS pipeline. Although acoustic techniques is widely used in the water industry for leakage detection and localization but it has several limitations. The limitations are it unable to work effectively in plastic pipeline because in the plastic pipeline it absorb the sound energy resulting in weaken the sound wave. When the sound wave propagate along the pipeline, it increase the high-frequency noise and the process to analyze the high-frequency noise is complicated. The leak detection is also affected due to air presence inside the pipeline [38]. This paper extends the work [38] by adding second ADXL335 three-axis vibration sensors to detect the leakage and localize the leak in plastic ABS pipeline and improvising the existing leakage localization equation to improve the leak localization accuracy.

## 3. EXPERIMENTAL SETUP

### 3.1 Cross-correlation Technique

Figure 3.1 illustrates the sample of cross-correlation graphs. The y-axis refers to the measurement parameters investigated and the x-axis refers to time arrival or time delay. The time delay is denoted by a negative value. Examined the highest peak from y-axis and the time difference is calculated from the x-value to correspond to the highest peak. Equation 1 is the cross-correlation formula used to measure the leak location [15, 39-42].

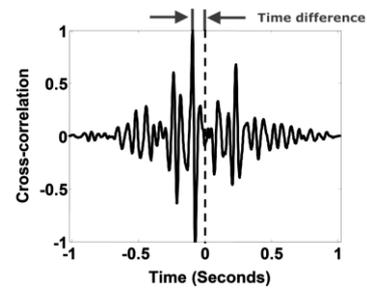

*Fig. 1. Cross-correlation graph.*

$$d_l = \frac{(L - c\Delta t)}{2} \quad (1)$$

$d_l$ = leak location
$L$ = distance between sensors
$c$ = wave speed
$\Delta t$ = time differences

### 3.2 Calculating The Wave Speed

Below is the procedure to calculate the wave speed
1. Measure the water flow.
   The unit is in liter per minute (*lpm*).
2. Convert the water flow from liter per minute (*lpm*) to liter per second (*lps*).
3. Convert the water flow from liter per second (*lps*) to meter cube per second ($m^3$/s). 1 liter is equivalent to 0.001 $m^3$.
4. From the datasheet, the ABS pipeline inner diameter is 0.0334 m. Measure the pipeline area (A) using equation 2.

Pipeline diameter $\quad D = 0.0334$ m $\quad (2)$
$$A = \pi r^2$$
$$= \pi \left(\frac{0.0334}{2}\right)^2$$
$$= 8.76 \times 10^{-4} \; m^2$$

5. To measure the wave speed used the Equation 3.
$$c = \frac{water\ flow}{pipeline\ area} \quad (3)$$
$$= \frac{m^3/s}{m^2}$$
$$= m/s$$

**Table 1** Wave speed and water pressure

| Pressure (kgf/cm²) | Water flow (*lpm*) | Wave speed (m/s) |
|---|---|---|
| 0.6 | 22.80 | 0.437 |
| 1.0 | 18.50 | 0.352 |
| 1.4 | 14.60 | 0.278 |

### 3.3 Experimental Setup

Figure 3.2 illustrated the experiment setup for water pipeline leakage localization testbed. The pipeline used is Acrylonitrile Butadiene Styrene (ABS) with the diameter of 10 inch and 10 meters length. The water pump function to pump the water from the water bucket. The water pressure is set into three different pressure and the pressure is 0.6 kgf/cm², 1.0 kgf/cm² and 1.4 kgf/cm². The valve labelled number 1 function to control the water pressure to follow the pressure set for this experiment. The pressure and flow meter to measure the water pressure and water flow. There are two vibration sensors located along the pipeline and between the leak locations. Vibration sensor 1 located on the right and vibration sensor 2 located on the left of the leak location. The leak location size is 1 mm. The green hose labelled number 10 function to circulate the water flow. The vibration data from the sensors transferred to the laptop through Zigbee wireless. In the laptop, a *'putty.exe'* application is used to monitor and record the data collected from the sensors.

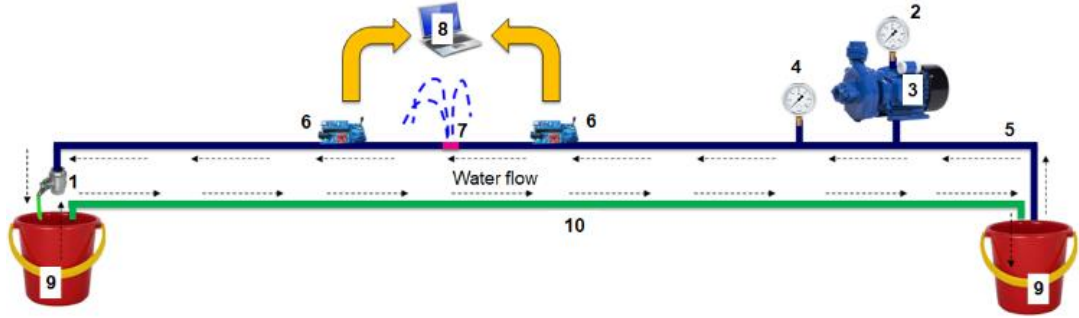

*Fig. 3.2.  Water pipeline leakage localization testbed*

**Legend:**

| | | | | | |
|---|---|---|---|---|---|
| 1 | Valve. | 5 | ABS water pipeline | 8 | Laptop (analysis) |
| 2 | Pressure meter | 6 | Vibration sensor and Zigbee module | 9 | Water bucket |
| 3 | Water pump | 7 | Leakage location | 10 | Hose |
| 4 | Flow meter | | | | |

Table 2  Experiment parameters

| No | Parameters | Values | Explanations |
|---|---|---|---|
| 1 | Leak size | 1 mm | Single leak drill on the marked location in the pipeline. |
| 2 | ADXL335 with Zigbee module | 2 units | Sensor 1 located on the left of the leak and sensor 2 located on the right of the leak equidistance each other. The vibration sensors will record the vibration data in three axis; x, y and z. |
| 3 | Interference | Motor, Valve, Water pump. | The vibration data collected under No Leak and Leak 1 mm condition will be filtered from the interference. The vibration sensor is placed at the three locations on the pipeline and data were collected under different water pressure. |
| 4 | Sensor distance | 0.5 m 1.0 m, 1.5 m, 2.0 m | This is the actual distance from the vibration sensors to leak location. Each sensor distance is tested with different water pressure and the vibration data is recorded from the three axis. The vibration data is processed and applied with cross-correlation techniques to calculate the leakage location. |
| 5 | Water pressure | 0.6 kgf/cm$^2$, 1.0 kgf/cm$^2$, 1.4 kgf/cm$^2$ | The water pressure is controlled by the valve.  Three different water pressures are applied throughout the experiment to analyze the vibration sensor performance to localize the leakages. |

All sets of data were collected for the duration of eight minutes and four sets of data. The experiment starts with collecting the vibration data under No Leak condition at the valve, water pump, and motor. The data were collected under three different water pressures.

Marked a leak location on the pipeline. Set the water pressure to 0.6 kgf/cm$^2$ and mounted the vibration sensors at a distance of 0.5 m to the left and right from the marked leak location. Record the vibration data from both sensors. Repeat this steps for the sensor distance of 1.0 m, 1.5 m and 2.0 m from the marked leak locations. Repeat this procedure again for different water pressure; 1.0 kgf/cm$^2$ and 1.4 kgf/cm$^2$.

Drill a hole size of 1 mm at the marked locations. Repeat the previous steps explain to collect the vibration data under different water pressure and different sensor distance.

Figure 2, 3 and 4 shows the cross correlation graph under No Leak and Leak 1 mm scenario. The cross correlation graph is from the two vibration sensors located in the right and left of the leak. The water flow from sensor 1 located on right of the leak to sensor 2 located on the left side of the leak. The x-axis corresponds to value of time and the y-axis corresponds to the vibration magnitude. The highest peak indicates a similar vibration signal observed in both sensors at time *t*. Negative *t* means there is a time delay for the similar vibration signal from sensor 1 to arrive in sensor 2 whilst positive *t* indicates time arrival. In No Leak scenario, the highest peak is located at x-axis equals or near to zero value as illustrated in figure 2(a), 3(a) and 4(a). In Leak 1 mm scenario, the highest peak from the cross correlation is located either to the left or right of the x-axis origin as illustrated in figure 2(b), 3(b) and 4(b).

## 4. RESULTS AND DISCUSSION

### 4.1  Leak Detection

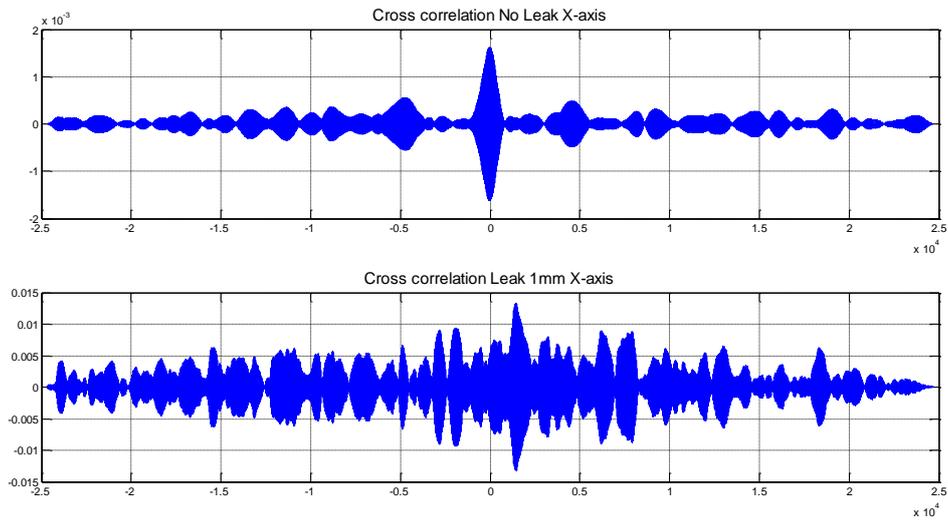

***Fig. 2.*** *Cross-correlation graph x-axis for sensor distance of 0.5 m and water pressure 0.6 kgf/cm$^2$. (a) No Leak and (b) Leak 1 mm*

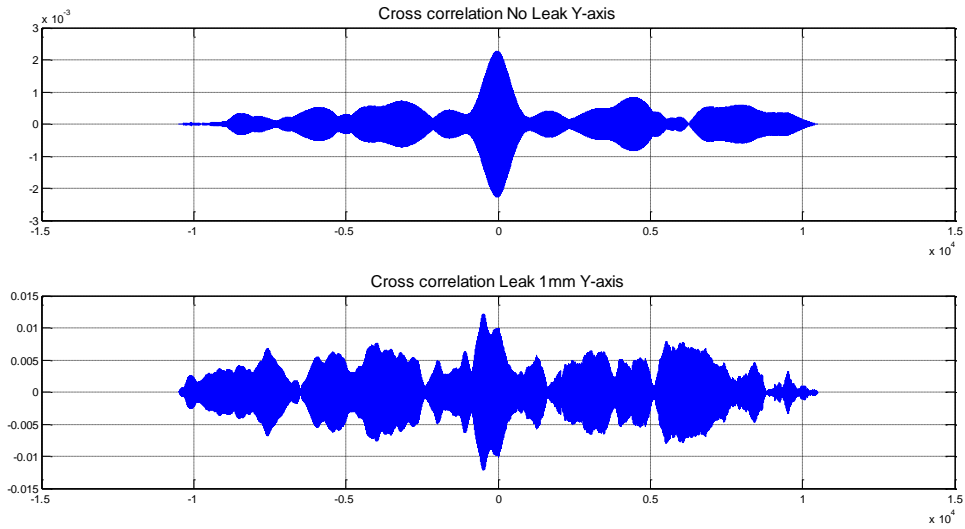

***Fig. 3.*** *Cross-correlation graph y-axis for sensor distance of 0.5 m and water pressure 1.0 kgf/cm$^2$. (a) No Leak and (b) Leak 1 mm*

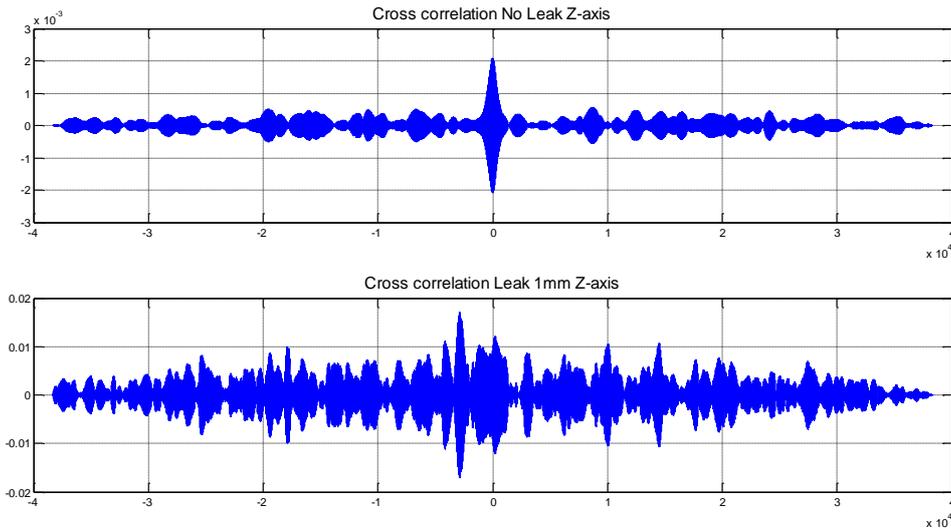

***Fig. 4.*** *Cross-correlation graph z-axis for sensor distance of 1.0 m and water pressure 1.4 kgf/cm$^2$. (a) No Leak and (b) Leak 1 mm*

From the cross correlation graph, it is easy to identify the pipeline condition just by observing the location of the highest peak value. If there is a leak in the pipeline, the highest peak is located either to the left or right of the origin of x-axis. Table 3 below presented all the recorded time from the cross correlation graph for pressure 0.6 kgf/cm$^2$, 1.0 kgf/cm$^2$ and 1.4 kgf/cm$^2$ for sensor distance 0.5 m, 1.0 m, 1.5 m and 2.0 m. Data in the row of 0.6(a), 1.0(a) and 1.4(a) ideally should be zero value instead of an integer value. However integer value was recorded because of the time differences to record the vibration data from two separate '*putty.exe*' application windows.

**Table 3** Time delay / time arrival from cross-correlation graph for pressure 0.6 kgf/cm$^2$, 1.0 kgf/cm$^2$ and 1.4 kgf/cm$^2$ for sensor distance 0.5 m, 1.0 m, 1.5 m and 2.0 m.

| Sensor Distance (m) | | 0.5 | | | 1.0 | | | 1.5 | | | 2.0 | | |
|---|---|---|---|---|---|---|---|---|---|---|---|---|---|
| | Axis | x | y | z | x | y | z | x | y | z | x | y | z |
| Water Pressure (kgf/cm$^2$) | 0.6 (a) | -0.14 | -0.14 | -0.14 | -0.12 | -0.12 | -0.12 | -0.14 | -0.14 | -0.14 | -0.14 | -0.14 | -0.14 |
| | 0.6 (b) | 14.49 | 13.97 | 79.00 | 22.09 | -28.20 | 20.81 | -39.01 | 3.62 | 3.58 | 12.25 | -218 | 1249 |
| | 1.0 (a) | -0.26 | -0.26 | -0.26 | -0.10 | -0.10 | -0.10 | -0.02 | -0.02 | -0.02 | -0.02 | -0.02 | -0.02 |
| | 1.0 (b) | 0.14 | -4.84 | -11.17 | 1.7 | 15.19 | 1.26 | -73.16 | -46.88 | -46.82 | 36.95 | 36.93 | 37.37 |
| | 1.4 (a) | -0.06 | -0.06 | -0.06 | -0.08 | -0.08 | -0.08 | -0.08 | -0.08 | -0.08 | -0.08 | -0.08 | -0.08 |
| | 1.4 (b) | 86.16 | -6.80 | 40.05 | -2.53 | -5.99 | -28.89 | -1.67 | -0.47 | 25.53 | -28.85 | -29.35 | -20.25 |

Data in the row of 0.6(b), 1.0(b) and 1.4(b) was recorded under Leak 1 mm condition. The integer value represented the time taken for the vibration signal in sensor 2 is identical with the vibration signal in sensor 1.

### 4.2 Leak Localization

From the cross correlation graph, the value of time is recorded. Equation 1 is used to calculate the leak distance from the sensor.

From Table 4 below, it is observed that, the leak accuracy is high in certain pressure and leak distance. Pressure 0.6 kgf/cm$^2$ and the actual sensor distance 2.0 m (y-axis). The calculated leak distance is 2.44 m and the error percentage is -22.00%

Pressure 1.0 kgf/cm$^2$ and the actual sensor distance of 0.5 m and 1 m. For an actual sensor distance of 0.5 m, the lowest error percentage is observed at x-axis (14.00%) which the calculated leak distance is 0.43 m. For an actual sensor distance of 1.0 m, the low error percentage is observed at x-axis (31.00%) and z-axis (23.00%). The calculated leak distance at x-axis is 0.69 and z-axis is 0.77 m.

Pressure 1.4 kgf/cm$^2$ and actual sensor distance of 1.0 m and 1.5 m. For an actual sensor distance of 1.0 m, the lowest error percentage is observed in x-axis (-34.00%) and the calculated leak distance is 1.34 m. For an actual sensor distance of 1.5 m, the lowest error percentage is at y-axis (-3.33%) and the calculated leak distance is 1.55 m.

While for the remaining water pressure and sensor distance, the localization accuracy is low. In the next section a time buffer ($t_{buffer}$) is proposed to improve the localization accuracy.

**Table 4** Leak distance calculated from cross correlation data

| Actual Sensor Distance (m) | | 0.5 | | | 1.0 | | | 1.5 | | | 2.0 | | |
|---|---|---|---|---|---|---|---|---|---|---|---|---|---|
| | Axis | x | y | z | x | y | z | x | y | z | x | y | z |
| Water Pressure (kgf/cm$^2$) | 0.6 | -2.67 | -2.56 | -16.66 | -3.82 | 7.09 | -3.54 | 3.52 | 7.70 | 5.75 | -0.69 | 2.44 | -0.74 |
| | Error (%) | 634.00 | 612.00 | 3432.00 | 482 | -609 | 454 | -134.67 | -413.33 | -283.33 | 134.50 | -22.00 | 137.00 |
| | 1.0 | 0.43 | 1.28 | 2.35 | 0.69 | -1.60 | 0.77 | 14.49 | 9.82 | 9.81 | -4.51 | -4.51 | -4.59 |
| | Error (%) | 14.00 | -156 | -370 | 31.00 | 260 | 23.00 | -866.00 | -554.67 | -554.00 | 325.50 | 325.50 | 329.50 |
| | 1.4 | -11.47 | 1.44 | -5.07 | 1.34 | 1.82 | 5.00 | 1.72 | 1.55 | -2.06 | 5.99 | 6.06 | 4.80 |
| | Error (%) | 2394 | -188 | 1114 | -34.00 | -82.00 | -400.00 | -14.67 | -3.33 | 237.33 | -199.50 | -203.00 | -140.00 |

## 4.3 Proposed Leakage Localization Equation

The proposed leak error accuracy is 2.9% [21]. In this experiment, the actual leak distance from the sensor are 0.5 m, 1.0 m, 1.5 m and 2.0 m. Table 5 below illustrated the ideal leakage distance based on error accuracy of 2.9%. The minimum value means the leak location calculated is before the actual leak location and the maximum value means the leak location calculated is after the actual leak location. Figure 5 ,6 and 7 shows a few of selected cross-correlation graphs under No Leak and Leak 1 mm conditions with marked *t*-value that corresponds to the minimum and maximum computed leak distance.

**Table 5** Computed ideal leak distance based on 2.9% error accuracy.

| Actual leak distance (m) | | 0.5 | 1.0 | 1.5 | 2.0 |
|---|---|---|---|---|---|
| Computed leak distance (m) | Min | 0.49 | 0.97 | 1.46 | 1.94 |
| | Max | 0.51 | 1.03 | 1.54 | 2.06 |

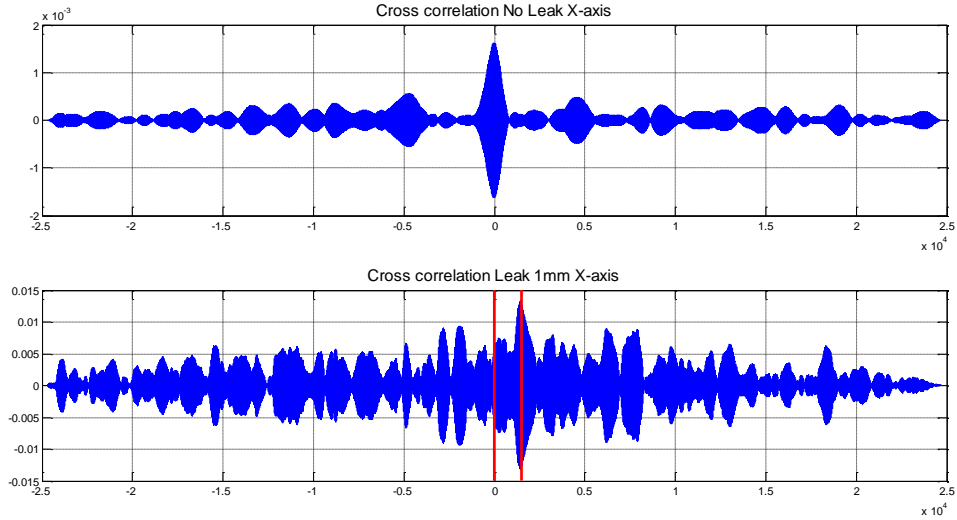

***Fig. 1.*** *Cross-correlation graph x-axis for sensor distance of 0.5 m and water pressure 0.6 kgf/cm$^2$. (a) No Leak and (b) Leak 1 mm with location of ideal t-value for acceptable leakage location calculation.*

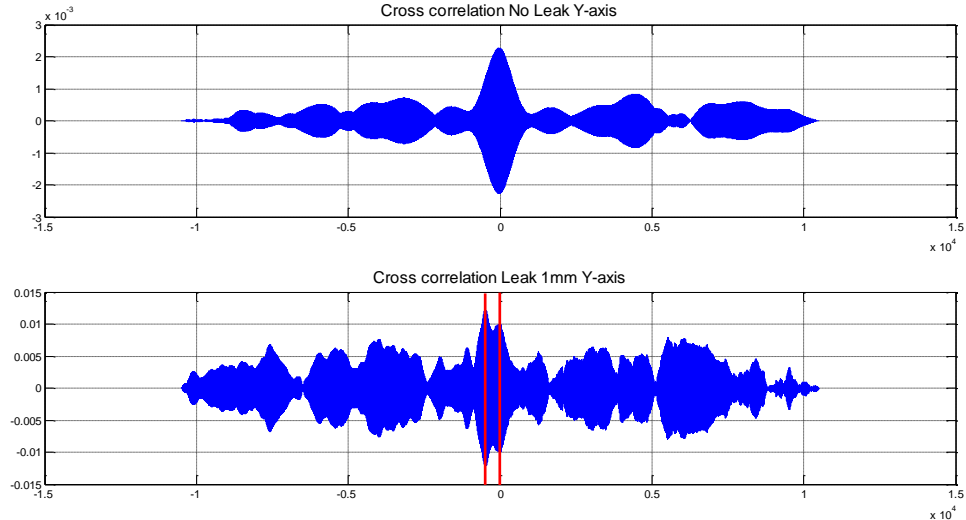

***Fig. 2.*** *Cross-correlation graph y-axis for sensor distance of 0.5 m and water pressure 1.0 kgf/cm2. (a) No Leak and (b) Leak 1 mm with location of ideal t-value for acceptable leakage location calculation.*

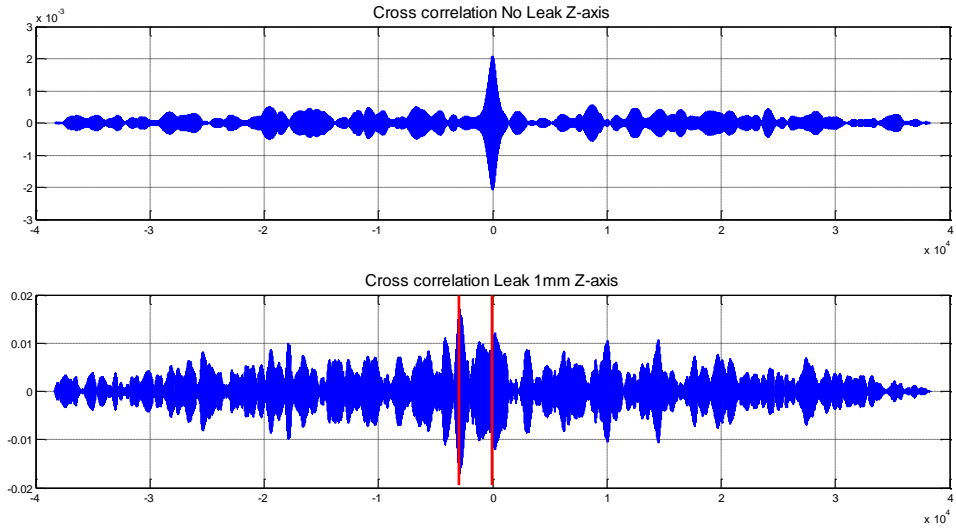

*Fig. 3. Cross-correlation graph z-axis for sensor distance of 1.0 m and water pressure 1.4 kgf/cm². (a) No Leak and (b) Leak 1 mm with location of ideal t-value for acceptable leakage location calculation*

In figure 5(b), 6(b) and 7(b), there are two vertical red lines. The two vertical red lines represents the ideal *t*-value for both minimum and maximum to obtain the acceptable calculated leak distance. The left vertical line is the minimum value and maximum value corresponds to the right vertical line.

Table 6 shows the *t* actual ($t_{actual}$) to achieve the recommended leak error accuracy. $\Delta t$ is the ideal *t*-value to obtained the acceptable calculated leak distance. $t_{noLeak}$ is the value of time under No Leak condition. $t_{actual}$ is calculated using equation 5 below.

$$\Delta t = t_{actual} - t_{noLeak} \quad (4)$$

Rearranging equation 4

$$t_{actual} = \Delta t + t_{noLeak} \quad (5)$$

To have an acceptable leak location, $t_{buffer}$ is introduced into the equation 6.

$$t_{ideal} = t_{actual} - t_{noLeak} \quad (6)$$

$$t_{ideal} = t_{actual} - t_{noLeak} - t_{buffer} \quad (7)$$

**Table 6** $t_{(ideal)}$ to achieve recommended leakage accuracy

| Pressure (kgf/cm²) | Sensor distance (m) | 0.5 | | 1.0 | | 1.5 | | 2.0 | |
|---|---|---|---|---|---|---|---|---|---|
| | | min | max | min | max | min | max | min | max |
| | $D_{l(ideal)}$ (m) | **0.49** | **0.51** | **0.97** | **1.03** | **1.46** | **1.54** | **1.94** | **2.06** |
| 0.6 | $\Delta t$ | 0.07 | -0.07 | 0.13 | -0.13 | 0.20 | -0.20 | 0.27 | -0.27 |
| | $t_{noLeak}$ | -0.14 | -0.14 | -0.12 | -0.12 | -0.14 | -0.14 | -0.14 | -0.14 |
| | $t_{actual}$ | -0.07 | -0.21 | 0.01 | -0.25 | 0.06 | -0.34 | 0.13 | -0.41 |
| 1.0 | $\Delta t$ | 0.09 | -0.09 | 0.17 | -0.17 | 0.24 | -0.24 | 0.33 | -0.33 |
| | $t_{noLeak}$ | -0.14 | -0.14 | -0.10 | -0.10 | -0.02 | -0.02 | -0.02 | -0.02 |
| | $t_{actual}$ | -0.05 | -0.23 | 0.07 | -0.27 | 0.22 | -0.26 | 0.31 | -0.35 |
| 1.4 | $\Delta t$ | 0.10 | -0.10 | 0.21 | -0.21 | 0.31 | -0.31 | 0.42 | -0.42 |
| | $t_{noLeak}$ | -0.06 | -0.06 | -0.08 | -0.08 | -0.08 | -0.08 | -0.08 | -0.08 |
| | $t_{actual}$ | 0.04 | -0.16 | 0.13 | -0.29 | 0.23 | -0.39 | 0.34 | -0.50 |

## 5. CONCLUSSION

Two vibration sensors can detect the leakage in ABS pipeline in all the three axis regardless of the leak distance and water pressure. Using the existing leakage localization formula, the ADXL335 vibration sensor only provide high accuracy of calculated leak location for certain water pressure, sensor distance and limited axis. Therefore an improvised equation is proposed by adding a time buffer ($t_{buffer}$) to improve the calculated leak location. $t_{buffer}$ is introduced into the equation and it improved the calculated leak location significantly for all three water pressures, all sensor distance and all three axis.

## ACKNOWLEDGEMENT

This work is supported by the Universiti Teknologi Malaysia under PAS, with cost center no Q.K130000.2740.00K70. We would also like to thank the Ministry of Higher Education and High Center of Excellence Wireless Communication Center (WCC) for funding the publication of the paper under R.K130000.7840.4J235. We would like to extend our gratitude to U-BAN members for comments on the work.